\documentclass[fleqn,usenatbib]{mnras}

\usepackage{newtxtext,newtxmath}

\usepackage[T1]{fontenc}
\usepackage{ae,aecompl}

\usepackage{graphicx}	
\usepackage{amsmath}	
\usepackage{amssymb}	

\title[LOFAR observations of GRB 180706A]{LOFAR early-time search for coherent radio emission from GRB 180706A}

\author[A. Rowlinson et al.]{A. Rowlinson$^{1,2}$\thanks{E-mail: b.a.rowlinson@uva.nl}, K. Gourdji$^1$, K. van der Meulen$^{1}$, Z.S. Meyers$^{1}$, T.W. Shimwell$^{2}$, \and S. ter Veen$^{2}$, R.A.M.J. Wijers$^{1}$, M.J. Kuiack$^{1}$, A. Shulevski$^{1}$, J.W. Broderick$^{3}$, \and A.J. van der Horst$^{4,5}$, C. Tasse$^{6,7}$, M.J. Hardcastle$^{8}$, A.P. Mechev$^{9}$, W.L. Williams$^{8}$\\
$^{1}$ Anton Pannekoek Institute, University of Amsterdam, Postbus 94249, 1090 GE Amsterdam, The Netherlands\\
$^{2}$ ASTRON, the Netherlands Institute for Radio Astronomy, Postbus 2, NL-7990 AA Dwingeloo, the Netherlands\\
$^{3}$ International Centre for Radio Astronomy Research, Curtin University, GPO Box U1987, Perth, WA 6845, Australia \\
$^{4}$ Department of Physics, the George Washington University, 725 21st Street NW, Washington, DC 20052, USA \\
$^{5}$ Astronomy, Physics and Statistics Institute of Sciences (APSIS), 725 21st Street NW, Washington, DC 20052, USA \\
$^{6}$ GEPI, Observatoire de Paris, Universite PSL, CNRS, 5 Place Jules Janssen, 92190 Meudon, France \\
$^{7}$ Department of Physics \& Electronics, Rhodes University, PO Box 94, Grahamstown, 6140, South Africa \\
$^{8}$ Centre for Astrophysics Research, School of Physics, Astronomy and Mathematics, University of Hertfordshire, \\ College Lane, Hatfield AL10 9AB, UK \\
$^{9}$ Leiden Observatory, Leiden University, PO Box 9513, NL-2300 RA Leiden, The Netherlands
}

\date{Accepted XXX. Received YYY; in original form ZZZ}

\pubyear{2015}

\begin{document}
\label{firstpage}
\pagerange{\pageref{firstpage}--\pageref{lastpage}}
\maketitle

\begin{abstract}
The nature of the central engines of gamma-ray bursts (GRBs) and the composition of their relativistic jets are still under debate. If the jets are Poynting flux dominated rather than baryon dominated, a coherent radio flare from magnetic re-connection events might be expected with the prompt gamma-ray emission. There are two competing models for the central engines of GRBs; a black hole or a newly formed milli-second magnetar. If the central engine is a magnetar it is predicted to produce coherent radio emission as persistent or flaring activity. In this paper, we present the deepest limits to date for this emission following LOFAR rapid response observations of GRB 180706A. No emission is detected to a 3$\sigma$ limit of 1.7 mJy beam$^{-1}$ at 144 MHz in a two-hour LOFAR observation starting 4.5 minutes after the gamma-ray trigger. A forced source extraction at the position of GRB 180706A provides a marginally positive (1 sigma) peak flux density of $1.1 \pm 0.9$ mJy. The data were time-sliced into different sets of snapshot durations to search for FRB like emission. No short duration emission was detected at the location of the GRB. We compare these results to theoretical models and discuss the implications of a non-detection.
\end{abstract}




\section{Introduction}

Since the first detection of gamma-ray bursts (GRBs), the community has been steadily gaining understanding of these events and their progenitor systems. Long GRBs are associated with core collapse supernovae \citep[e.g.][]{hjorth2003,woosley2006} and short GRBs occur following the merger of two neutron stars \citep[confirmed by the detection of GW 170817 and its association with GRB 170817A;][]{abbott2017} or a neutron star and a black hole. We now know that the prompt gamma-ray emission from GRBs can be accompanied by TeV gamma-ray emission \citep{mirzoyan2019}, X-rays and optical flashes. Astronomers have also searched for prompt radio emission, that could be associated with the central engine or the relativistic jet, but with no detections to date \citep{Cortiglioni1981,Inzani1982,Koranyi1995,Dessenne1996,Balsano1998}. These non detections are likely due to the small sample sizes of these studies (not all GRBs are likely to produce detectable radio emission in the same way as not all GRBs produce optical flashes) and relatively insensitive searches, typically $\gtrsim 100$ Jy. One survey found a tantalising hint of two Fast Radio Bursts (FRBs), though at very low significance, associated with the X-ray plateau phases of two long GRBs \citep{bannister2012} but this has yet to be confirmed. With the exception of the work by \cite{bannister2012}, which used rapid response observations by the Parkes Radio Telescope, the previous surveys have typically been either whole sky instruments (with limited sensitivity) or hampered by very slow slew times. Recently, astronomers used the Murchison Widefield Array \citep[MWA;][]{tingay2013}, a low frequency radio telescope array with no moving parts, to enable a very rapid response observation of a short GRB \citep{kaplan2015} reaching a sensitivity of $\sim$1 Jy on 30 minute time scales. Additionally, the Long Wavelength Array \citep[LWA;][]{taylor2012}, a whole sky transient survey instrument, was able to constrain prompt emission from a short GRB to a $1\sigma$ flux density limit of 4.5 Jy beam$^{-1}$ \citep{anderson2018}.

An origin of prompt coherent radio emission from GRB events could be from magnetic re-connection events within a Poynting flux dominated jet \citep[e.g.][]{drenkhahn2002}. The structure of the relativistic jets causing GRBs are still subject to investigation with Poynting flux dominated or baryon dominated jets being favoured \citep[e.g.][]{sironi2015}. We have observed coherent radio emission from magnetic re-connection events in the Sun, leading to a good understanding of the plasma physics involved \citep[e.g.][]{bastian1998}. Therefore, limits on coherent radio emission associated with the prompt emission can be used to constrain the presence of magnetic re-connection events \citep[e.g.][]{Inzani1982}. One such model has been proposed by \cite{usov2000}, who consider the coherent radio emission expected from relativistic, strongly magnetised winds produced by GRBs.

Additionally, the nature of the central engine powering these GRBs is a subject of continuing debate with two key theories proposed: a black hole \citep[e.g.][]{woosley1993} or a millisecond spin period, highly magnetised, massive neutron star \citep[hereafter referred to as a magnetar; e.g.][]{metzger2011}. An observable signature of the magnetar model is a prolonged X-ray plateau phase \citep{zhang2001}. As accretion ends within seconds for short GRBs, the plateau phases observed are typically associated with the magnetar model \citep[e.g.][ and references therein]{rowlinson2013}. However, for long GRBs, this plateau phase has been both associated with the magnetar central engine model \citep[e.g.][]{bernardini2012} and with ongoing accretion onto the central engine \citep{kumar2008}. Additional information will be required to more confidently associate these plateaus with the magnetar model. One of the predictions of the magnetar central engine model is the presence of coherent radio emission from the newly formed magnetar, associated with the plateau phase; this would not be present for the black hole central engine model. Thus, detection of persistent coherent radio emission during the plateau phase of a long GRB would likely rule out a black hole central engine. However, it remains unclear if this emission is able to escape from the local and galactic environment surrounding the GRB location \citep[e.g.][]{macquart2007,lyubarsky2008,zhang2014}. These plateau phases are expected to begin at the time of the GRB, though initially significantly lower flux than the prompt emission, they have durations ranging from 10 seconds up to a day (for the more extreme plateaus) and most plateaus end within a few hours  \citep[e.g. ][]{bernardini2012}. There are three key coherent emission models to test:
\begin{itemize}
  \item persistent pulsar-like emission from the magnetar engine \citep[e.g.][]{totani2013}.
  \item FRBs from the young, highly magnetised, neutron star \citep[e.g.][]{katz2016,lyutikov2016}
  \item a single FRB at the end of the plateau phase if the neutron star is too massive to support itself and it collapses to form a black hole \citep{falckerezzolla14,zhang2014}.
\end{itemize}

In November 2017, the LOw Frequency ARray \citep[LOFAR;][]{vanhaarlem2013} completed implementing a new rapid response mode, with observations using the full Dutch array starting within 5 minutes of receiving an alert\footnote{http://www.astron.nl/radio-observatory/lofar-system-capabilities/responsive-telescope/responsive-telescope}. Although this response time is slower than that of the MWA ($\sim$30 seconds), it is sufficiently fast to study the plateau phase and highly dispersed events. By utilising the full Dutch array, a large bandwidth and a two hour observation, we can attain the sensitivity required to deeply probe for emission during the plateau phase. We successfully requested a number of rapid response triggers on GRBs detected by the Niel Gehrels Swift Observatory \citep[hereafter referred to as {\it Swift};][]{gehrels2004} and, on 2018 July 6, we successfully completed our first, fully automated, rapid response trigger on GRB 180706A \citep{stamatikos2018}.

This paper will describe the constraints we can make using the LOFAR observations of GRB 180706A. In Section \ref{sec:grb180706A}, we describe the observations of GRB 180706A obtained, using {\it Swift} and the rapid response mode of LOFAR, and outline our analysis of these data. In Section \ref{sec:predictions}, we consider general coherent radio properties of the emission that can be constrained, we then compare our observations to theoretical models for coherent radio emission associated with both the relativistic jet and the central engine.

\section{Observations of GRB 180706A}
\label{sec:grb180706A}

\subsection{{\it Swift} Observations}
GRB 180706A was detected by the Burst Alert Telescope \citep[BAT;][]{barthelmy2005} on board the {\it Swift} Satellite at 08:24:40 UT on 2018 July 6 \citep{stamatikos2018}. This GRB was also independently detected by the {\it Fermi} Gamma-ray Burst Monitor \citep[GBM;][]{meegan2009, bissaldi2018}. With a $T_{90}$ duration of $42.7\pm8.7$ seconds (15--350 keV), this is a long GRB with a likely collapsar progenitor \citep{woosley1993}. 

{\it Swift} automatically slewed to the position of GRB 180706A and X-ray Telescope \citep[XRT;][]{burrows2005} observations started 87.7 seconds after the trigger \citep{stamatikos2018} and a bright and rapidly fading X-ray counterpart was observed. At 95 seconds after the trigger, the UV and Optical Telescope on board {\it Swift} \citep[UVOT;][]{roming2005} started conducting observations and detected the optical counterpart at $19.88\pm0.34$ mag \citep[white filter;][]{oates2018}. This optical counterpart was also detected by other facilities and shown to be fading \citep{watson2018,volnova2018,ulaczyk2018}.

The gamma-ray data from the BAT and the unabsorbed X-ray data from XRT were obtained using the {\it Swift} Burst Analyser \citep{evans2010} in the 0.3--10 keV energy band. The light curve is characterised by three bright peaks followed by a steep decay and a plateau phase. At $\sim 10^{4}$ s, the plateau turns over to a power law decay phase. Unfortunately, no redshift was obtained for GRB 180706A, though we infer an upper limit of $z\lesssim2$ from the lowest wavelength in which the optical afterglow was detected \citep{gcnUVOT}. 

\subsection{LOFAR Observations}
\label{sec:LOFARobs}

\begin{figure}
\centering
\includegraphics[width=0.48\textwidth]{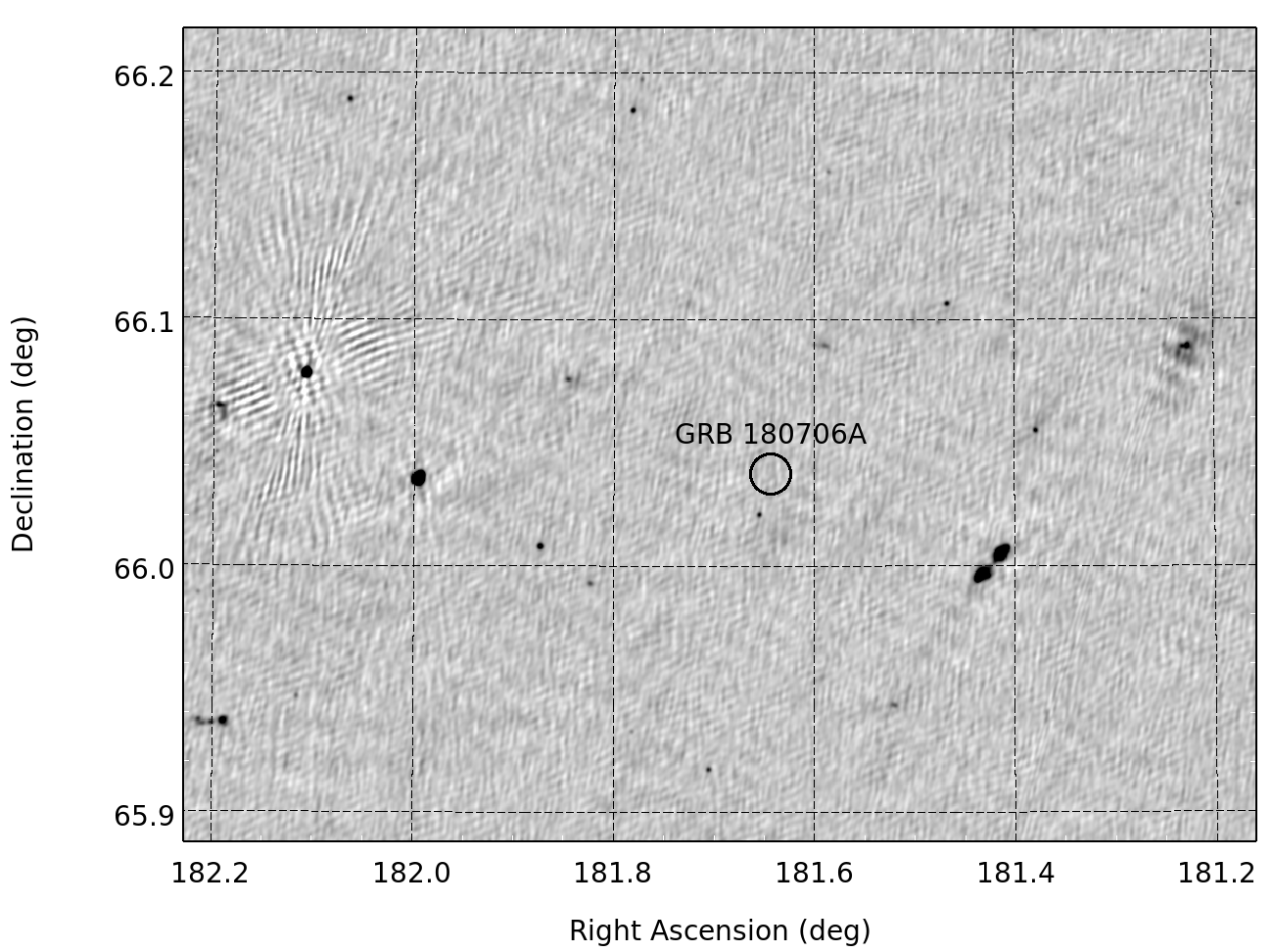}
\caption{This 2 hour integration LOFAR image shows the region surrounding GRB 180706A at 144 MHz. The circle shows the position of GRB 180706A and there is a 3$\sigma$ upper limit of 1.7 mJy beam$^{-1}$ for the flux density of this event.}
\label{fig:2hrImg}
\end{figure}

We received the VOEvent \citep{williams2006} published by {\it Swift} via the {\sc 4 Pi Sky broker} \citep[using {\sc Comet}; ][]{staley2016,swinbank2014} and, after checking the GRB met our criteria (including source elevation and calibrator availability), automatically triggered observations by sending an {\sc xml} observing request to the LOFAR system. Our trigger was successfully scheduled and a two hour observation started at 08:29:14 UTC on 2018 July 6, approximately 4.5 minutes after the GRB. We used the BAT position as the pointing centre of our observation. Immediately following this observation, we automatically completed a 2 minute observation of the calibrator source, 3C295.

Our observational set-up was chosen to closely match that of the LOFAR Two-meter Sky Survey \citep[LoTSS;][]{shimwell2018}; this choice allows us to utilise the deep 8-hour survey images, when available, for comparison to our observations on the event of a detection of radio emission at the GRB location. The observations were completed using the LOFAR High Band Antennas (HBA), with a frequency range of 120--168 MHz and a central frequency of 144 MHz, covered by 244 sub-bands each with a bandwidth of 195.3 kHz. We used only the Dutch LOFAR stations, 22 core stations and 11 remote stations, covering projected baselines of 24 m to 60 km. The data were recorded using a time-step of 1 second and 64 channels per sub-band. We used the standard methods to pre-process our observations \citep{vanhaarlem2013}, keeping the 1 second time-step in the archived observations but averaging to 16 frequency channels per sub-band to reduce data volume.  

\subsubsection{Calibration}

We used \textsc{prefactor}\footnote{https://github.com/lofar-astron/prefactor} to calibrate the target data, following and adapting the strategy used in \cite{vanweeren16factor} accordingly. This processing includes flagging of baselines with excess radio frequency interference (RFI) using the {\sc AOFlagger} \citep{offringa2010,offringa2012}. Additionally, baselines containing the stations CS021 and CS026 were flagged due to increased noise from these stations. Finally, contributions from the brightest radio sources in the sky, referred to as the A-team, were flagged. For this analysis, the calibrator and target data were both averaged in time to 10\,s and 48.82\,KHz (4 channels per subband). Diagonal gain solutions were obtained toward the calibrator source, 3C295, using the model defined by \cite{scaife2012}. 

The calibrator gain solutions were transferred to the target visibility data. The target subbands were combined in groups of 27, resulting in combined datasets of 5.272\,MHz. We obtained a sky model of the target field using the global sky model developed by \cite{scheers2011} and the TIFR GMRT Sky Survey at 150 MHz \citep[TGSS; ][]{intema2017}\footnote{TGSS catalogue: http://tgssadr.strw.leidenuniv.nl/doku.php}. Phase calibration of these datasets was carried out on a 10\,s time scale, using this skymap of the target field.

\subsubsection{Imaging}
\label{sec:imaging}

We used the LoTSS pipeline\footnote{\url{https://github.com/mhardcastle/ddf-pipeline}} in the manner described by \cite{shimwell2018} to carry out direction-dependent self-calibration and imaging of the full 2-hour observation. The final product was a direction-dependent calibrated image of the full dataset, made using the direction-dependent imager {\sc DDFacet} \citep{tasse2018}, with a central frequency of 144 MHz and a bandwidth of 48 MHz, using the settings outlined in \cite{shimwell2018}. We show the region surrounding GRB 180706A in Figure \ref{fig:2hrImg}. The image RMS at GRB location (30 arcsec radius) is 0.58 mJy beam$^{-1}$, corresponding to a 3$\sigma$ upper limit of 1.7 mJy beam$^{-1}$. Using the Python Source Extractor \citep[{\sc PySE};][]{carbone2018} we also conduct a forced source extraction at the position of the GRB holding the shape and size of the Gaussian shape fitted fixed to the restoring beam shape. We measure a peak flux density of $1.1 \pm 0.9$ mJy beam$^{-1}$ \citep[the uncertainty on this value is as measured by {\sc PySE} and hence excluding the image RMS noise][]{carbone2018}.

In addition to the 2 hour integrated image, we also imaged the target data on four snapshot time scales to search for short duration coherent radio emission. We used the sources modelled in the 2 hour integrated image, during the direction dependent and self calibration stages, to subtract them from the target visibilities. This subtraction enables us to probe deeper at the location of the GRB.

We created Stokes I images of these source-subtracted visibilities using {\sc WSClean} \citep{offringa2014}\footnote{http://wsclean.sourceforge.net} with Briggs weighting, a pixel scale of 10 arcseconds and baselines up to 12 km. Using the intervals-out option in {\sc WSClean}, we created snapshot images of durations 30 seconds, 2 minutes, 5 minutes and 10 minutes (the motivation for this range of time scales is outlined in Section \ref{sec:FRB}). The images have a typical angular resolution of $\sim$30 arcsec.

\begin{figure}
\centering
\includegraphics[width=0.45\textwidth]{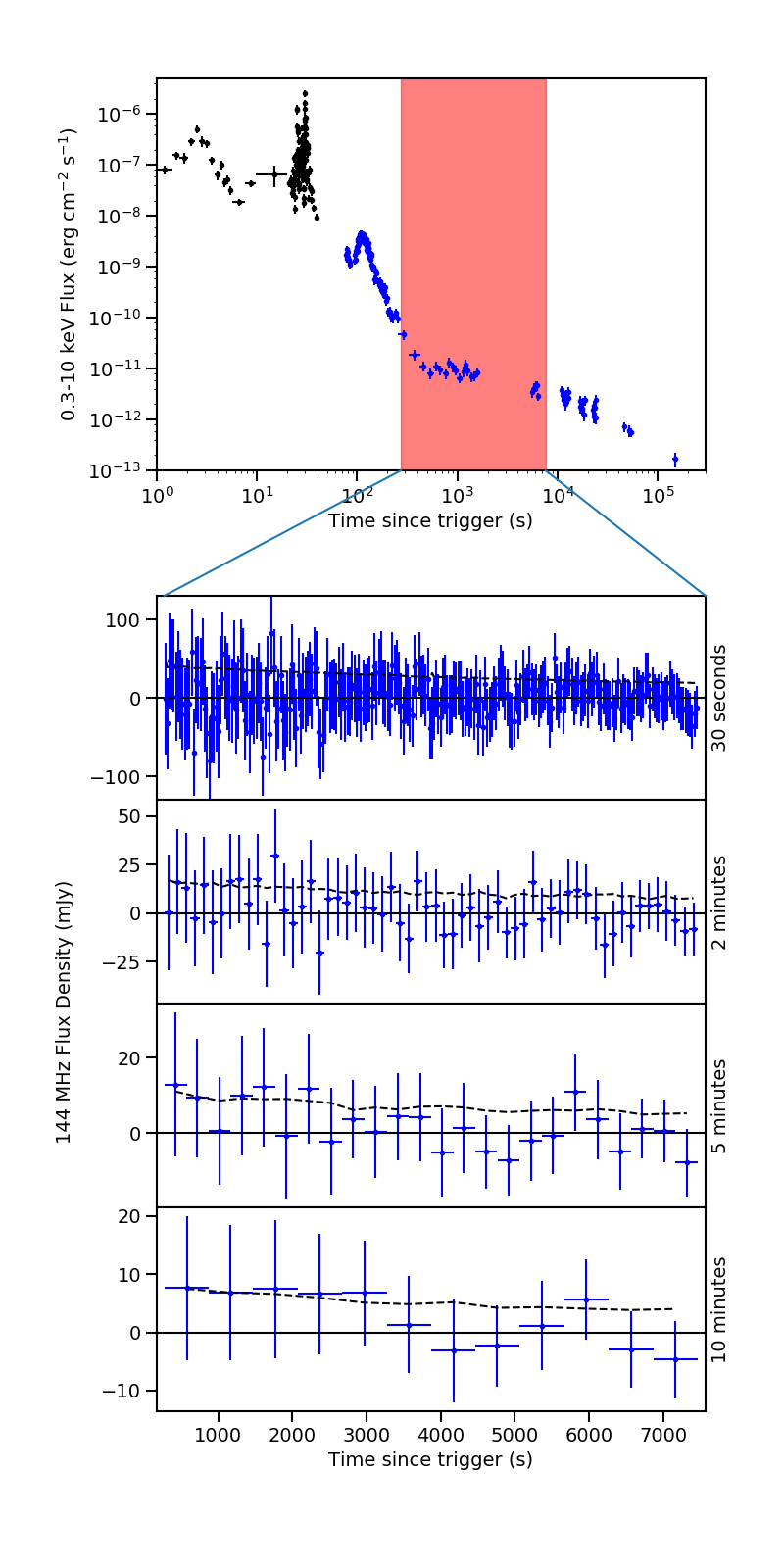}
\caption{In the top panel of this figure we show the 0.3--10 keV flux light curve of GRB 180706A. The black data points were obtained by the BAT (extrapolated to 0.3--10 keV) and the blue data points are from the XRT (observed at 0.3--10 keV). The red shaded region illustrates the time of the LOFAR observation. In the bottom panel, we plot the 144 MHz radio flux density observations as a function of time since the GRB trigger obtained by LOFAR. We show four different snapshot time scales: 30 seconds, 2 minutes, 5 minutes and 10 minutes. The solid black line shows a flux density of 0 mJy and the black dashed lines in each of the LOFAR light curves are the rms noise of the images; measured from the inner $\frac{1}{8}$th of the image.}
\label{fig:observedLC}
\end{figure}

\subsubsection{Image analysis using the LOFAR Transients Pipeline}

The images created were input into the LOFAR Transients Pipeline \citep[{\sc TraP}; ][]{swinbank2015}\footnote{https://github.com/transientskp/tkp}; a pipeline designed to automatically process radio images to search for transient and variable sources. This pipeline uses {\sc PySE}, the sourcefinder also used in Section 2.2.2. This pipeline measures the rms (root mean square) noise in the inner $\frac{1}{8}$th of the images. In Table \ref{table:rms}, we give the typical rms noise for the different imaging time scales.

\begin{table}
\centering
\begin{tabular}{|c c|} 
\hline
Time scale & rms noise \\
(minutes)  & (mJy beam$^{-1}$) \\
\hline
0.5        & $ 28\pm 6 $ \\
2          & $ 11\pm 2 $ \\
5          & $ 7\pm 2 $ \\
10         & $ 5\pm 1 $ \\
\hline
\end{tabular}
\caption{The average rms noise for the images from each time scale with the 1$\sigma$ standard deviation.}
\label{table:rms}
\end{table}

We use the monitoring list capability of {\sc TraP}. Inputting the GRB location into the monitoring list option causes {\sc TraP} to measure the flux density at the location of the GRB\footnote{For further information see \cite{swinbank2015} and the {\sc TraP} documentation; http://docs.transientskp.org}.

In Figure \ref{fig:observedLC}, we show the observations obtained by {\it Swift} and the light curves produced by {\sc TraP} for each of the different time scales of snapshot images (the integration time of each image is shown by the horizontal error bars), with the image rms over plotted with the black lines. As can be seen from this Figure, the flux densities at the GRB location are consistent with the noise in the inner $\frac{1}{8}$th of the images (n.b. in some images the local flux measurement, represented by the blue data points, can be lower than the image rms due to the local rms being slightly lower than the image rms). We note from these snapshot images that the observed flux density at the location of the GRB is marginally positive but consistent with zero within the uncertainties. Therefore, no coherent emission was detected from GRB 180706A in this analysis.

\section{Theoretical interpretation of observations}
\label{sec:predictions}

In the previous Section, we showed that we did not detect any coherent radio emission from GRB 180706A to deep limits. In this Section, we compare our non-detection to the theoretical models that predicted emission during the plateau phase. 

\subsection{Propagation considerations}
\label{sec:propagation}

First, we consider if coherent radio emission is able to escape the dense region surrounding the central engine and the surrounding host galaxy. \cite{macquart2007} showed that induced Compton and Raman scattering can significantly impede the passage of the coherent radio emission. They show that the emission can only escape if it is ultra relativistic ($\Gamma \gtrsim 10^3 \frac{D}{100 ~{\rm Mpc}}$) or emitted into a very narrow cone. To date, only lower limits on the GRB Lorentz factors have been observed \citep[e.g.][]{Ackermann2010,zhao2011} and they are known to be narrowly beamed although precise jet opening angles are still being constrained; it is therefore unknown if this is sufficient for the emission to escape. \cite{zhang2014} showed that the emission is likely to escape in the case of short GRBs. Therefore, it is unclear if the emission can escape but, if it does, it would place constraints on the Lorentz factor and the jet opening angle.

Once the coherent radio emission has escaped the immediate surroundings of the GRB, it still needs to travel through the host galaxy and this can lead to further absorption and scattering. Long GRBs are typically found in dense star formation regions near the centres of their host galaxies, making it likely that there is a large absorption column between the event and the Earth. Using the observed absorption in the X-ray spectrum, we can gain an understanding of the total absorption that the coherent radiation will pass through. The X-ray spectrum of GRB 180706A is best fit by a power law, with a photon index of $2.16^{+0.17}_{-0.16}$, and a total absorption column of $N_H = 6.7^{+3.7}_{-3.3} \times 10^{20}$ cm$^{-2}$ \citep[including the Galactic component from the Milky Way of $6.7 \times 10^{20}$ cm$^{-2}$;][]{Willingale2013}. Therefore, there are relatively low levels of absorption between the GRB location and {\it Swift} so the GRB likely occurred away from the most dense regions in its host galaxy increasing the likelihood of getting prompt radio emission to be observable once it has escaped from the central engine.

Therefore, while it looks more promising for short GRBs, we think there is a chance that the emission from the central engine would have been able to propagate to the Earth for this long GRB.

\subsection{Constraints on Fast Radio Bursts}
\label{sec:FRB}

There is evidence that repeating FRBs may be linked to Long GRBs. The host galaxy type of the only localized repeating FRB, FRB~121102, is the preferred environment of Long GRBs \citep{tendulkar17,marcote17,bassa17,kokubo17}. This is consistent with evolutionary links between FRBs and long GRBs, and possibly between FRBs and magnetars by extension. In this scenario, FRBs are produced from a young magnetar embedded in a supernova remnant \citep[SNR, e.g.][]{metzger17}. Using FRB~121102 as a prototype, \cite{law17} estimate the volumetric rate of FRBs and find it to be consistent with the rate of Long GRBs. \cite{nicholl17} also reach a similar conclusion, supportive of the Long GRB connection to FRBs. A potential caveat with detecting this type of emission very soon after the Long GRB is that the ejecta surrounding the magnetar may prevent the FRBs from escaping causing them to be be detectable only once the SNR has sufficiently expanded. For instance, \citet{cao17} and \citet{metzger17} estimate a minimum age of $\sim$10-100 years for FRB~121102. Indeed, other FRB sources detected in real time have not led to counterpart detections at other frequencies \citep{Petroff15,Keane16,Ravi16,Petroff17,Bhandari18,Farah19}. However, in this study, we are searching for coherent emission before a SNR has had the time to materialize and assume in the same way as in the previous discussion that the FRB emission can escape. Previous prompt FRB searches following Long GRB detections did not result in a firm detection \citep{bannister2012,Palaniswamy14}.

Another possible way to obtain an FRB is via the collapse of the central engine into a black hole as it becomes too massive to support itself \citep{falckerezzolla14,zhang2014}, which would be expected at the end of the plateau phase.

It is unclear whether FRBs are detectable at 144\,MHz. The lowest frequency at which FRBs have been detected is 400\,MHz (the bottom of the CHIME/FRB observing band) \citep{Chime19}. However, the fact that some of the FRBs detected at 400\,MHz show no or negligible scattering, suggests that FRBs are detectable at lower frequencies. The dearth of FRBs detected at lower frequencies may be due to the intrinsic FRB emission mechanism. More likely, though, is free-free absorption
at low frequencies \citep[e.g.][]{lyutikov2016,piro16}. At cosmological distances, however, the turnover in the rest frame spectrum is Doppler shifted \citep{Rajwade17}:
\begin{eqnarray}
\label{eqn:dopplerturnover}
\nu_{\rm rest} = \nu_{\rm observed}(1+z).
\end{eqnarray}
Using a rest frame spectral turnover due to free-free absorption of 300\,MHz \citep{lyutikov2016} and setting $\nu_{\rm observed}=120$\,MHz (the bottom of our observing band), our observations could be sensitive to long GRBs at $z>1.5$. We have assumed in this discussion that the plasma frequency, $\nu_p$ is below 144\,MHz and therefore is not a limiting factor (for $\nu_p>144$\,MHz, the electron number density would have to be at least $1.6\times10^{4}$\,cm$^{-3}$).

Assuming that the FRB emission is able to escape, we may be able to detect it in our snapshot images. To optimise the chances of detection, here we calculate a range of snapshot durations that probe different minimum flux densities and dispersion measure (DM) regimes. As the snapshot durations are greater than the width of the signal, we can follow the method described by \cite{trott2013} and estimate the minimum FRB flux densities $S_{\rm FRB,min}$ that we are sensitive to, using
\begin{eqnarray}
S_{\rm FRB,min} = S_{\rm img} \left( \frac{\Delta t_{\rm int}}{w} \right), \label{eqn:limFlx}
\end{eqnarray}
where $S_{\rm img}$ is the sensitivity in one snapshot image, $\Delta t_{\rm int}$  is the snapshot integration time and $w$ is the intrinsic width of the FRB. For consistency with previous works, we assume that the intrinsic width is 1 ms. 

\begin{figure}
\centering
\includegraphics[width=0.5\textwidth]{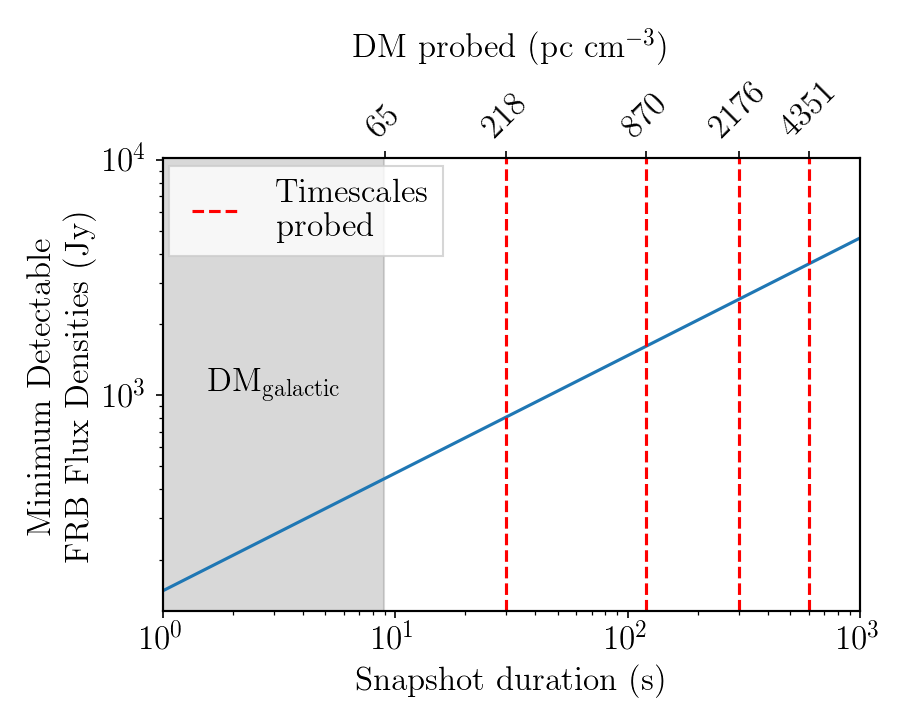}
\caption{Minimum FRB flux density that can be detected in our 144-MHz LOFAR data as a function of snapshot duration, assuming the duration of the dispersed signal is equivalent to the snapshot duration. The snapshot durations used in this study are marked by red dashed vertical lines. The top x-axis shows the corresponding DM value. The shaded region covers the galactic DM contribution (DM$_{\rm galactic}=~65\,\rm{pc\,cm^{-3}}$; see Section \ref{sec:FRB}) to the total DM along the line of sight of GRB~180706A. This acts as a lower limit in our search parameter space.
}
\label{fig:snapshots}
\end{figure}

Image noise scales with integration time as 
\begin{eqnarray}
S_{\rm img}\propto \Delta t_{\rm int}^{-\frac{1}{2}} \, .
\label{eqn:scale}
\end{eqnarray}
We apply $S_{\rm img}$=1.7\,mJy beam$^{-1}$ of the 2-hour observation to Equation \ref{eqn:scale} in order to obtain a scaling relationship for our data, and substitute this into Equation \ref{eqn:limFlx} to create the relationship between the minimum FRB flux to which we are sensitive as a function of snapshot duration, shown in Figure \ref{fig:snapshots}.

Equation \ref{eqn:limFlx} is true only if the duration of the dispersed signal, $\Delta t_{\rm dispersion}$, is equal to $\Delta t_{\rm int}$. The dispersion delay of a radio signal is described as \citep{pulsarhandbook}:
\begin{eqnarray}
\Delta t_{\rm dispersion} = 4.15\times10^3 \rm{MHz}^2 \rm{pc}^{-1} \rm{cm}^3 \rm{s}\bigg(\frac{1}{\nu_{\rm bottom}^{2}}-\frac{1}{\nu_{\rm top}^{2}}\bigg) DM \, ,
\label{eqn:dispersion}
\end{eqnarray}
where DM is the dispersion measure expressed in pc\,cm$^{-3}$ (i.e. the integrated column density of free electrons along the line of sight), and $\nu_{\rm bottom}=120.5$\,MHz and $\nu_{\rm top}=167$\,MHz are the limits of the observing band. In the top x-axis of Figure \ref{fig:snapshots}, we have substituted $\Delta t_{\rm int}$ into Equation \ref{eqn:dispersion} to visualize the optimal snapshot duration for a given FRB's DM value.

Thus, the choice of snapshot time scales to use is an optimization between DM probed and FRB sensitivity. We can set a minimum snapshot duration of $\sim$9\,s, which corresponds to the galactic DM contribution along the line of sight of GRB~180706A, DM$_{\rm Galactic}=~65\,\rm{pc\,cm^{-3}}$. We use four snapshot time scales in this study: 30 seconds, 2 minutes, 5 minutes, and 10 minutes (as described in Section \ref{sec:imaging}). Therefore, this experiment is sensitive to FRBs with flux densities in excess of 800\,Jy (value extrapolated from Figure \ref{fig:snapshots}). 
Using a reasonable FRB flux density of 1\,Jy at 1.4\,GHz, our experiment sensitivity would require a spectral index $\alpha\lesssim-3$. We note that FRB fluxes at $\sim1.4$\,GHz have reached $\sim150$\,Jy, which would flatten the spectral index required to $\sim-0.8$. 




Above we have demonstrated the important role of source DM in searches for coherent emission in radio images. It is difficult to predict the DM of coherent radio emission associated with GRB180706A. The main components of the total DM are the local burst environment (DM\textsubscript{local}), the host galaxy (DM\textsubscript{host}), the intergalactic medium (DM\textsubscript{IGM}), the Milky Way's halo (DM\textsubscript{halo}) and the interstellar medium (DM\textsubscript{ISM}, i.e. from the Milky Way's disk and spiral arms). According to the NE2001 model of free electrons in our Galaxy \citep{ne2001}, DM\textsubscript{ISM} should be $35\pm7$\,pc\,cm$^{-3}$ along the line of sight of GRB180706A. Following \citet{dolag2015}, DM\textsubscript{halo} is 30\,pc\,cm$^{-3}$. Using an assumed redshift $z=0.2$ allows us to follow the line of thinking from \citet{tendulkar17} for FRB~121102, which has $z\approx0.19$ to estimate a mean DM contribution from the IGM DM\textsubscript{IGM}$\approx200\pm85$\,pc\,cm$^{-3}$. It is difficult to estimate DM\textsubscript{host} without a host galaxy identification, as the value depends on galaxy type and viewing angle, though previous FRB related studies have used 100 pc\,cm$^{-3}$ \citep[e.g.][]{thornton13,XuHan15,caleb16}. As a lower limit at $z=0.2$, we estimate $\textrm{DM\textsubscript{total}}\gtrsim243 \rm{pc\,cm^{-3}}$.
In the upper limit case of $z\sim2$, DM\textsubscript{IGM} dominates over both DM\textsubscript{host} and DM\textsubscript{galactic} with values that can reach 2000\,pc\,cm$^{-3}$ \citep{ioka2003,inoue2004}. 
The DM\textsubscript{local} component (i.e. from the SNR) is difficult to estimate but can be significant \citep{connor2016}, particularly in the first few years of the neutron star.

Clearly, estimating the source DM beyond the contributions from our Galaxy is difficult and convoluted. Cases where the GRB has an associated host galaxy facilitate constraints on the total DM estimation. A larger population of FRBs should eventually provide insight as to the distribution of free electrons in the IGM. Until then, a more targeted search for coherent emission in radio images can be accomplished through image-plane de-dispersion. However, this is beyond the scope of the paper and will be presented elsewhere. In this study, we have simply chosen timescales that take into account the estimated minimum DM value and that probe parts of the search parameter space (Figure \ref{fig:snapshots}) within reason given the uncertainties involved.



\subsection{Constraints on prompt coherent radio emission}

If there were a radio flare emitted from the same region as the gamma-ray emission, it would be reasonable to assume that they could originate from the same emission mechanism. Taking the model by \cite{usov2000}, we assume that the gamma-ray and coherent radio emission both originate from magnetic re-connection in strongly magnetized winds within the GRB relativistic jet. In this scenario, the bolometric radio fluence, $\Phi_r$, is directly proportional to the bolometric gamma-ray fluence, $\Phi_{\gamma}$, where the power ratio $\langle \delta \rangle$ is given by:
\begin{eqnarray}
    \langle \delta \rangle = \frac{\Phi_r}{\Phi_{\gamma}}. \label{eqn:powerRat}
\end{eqnarray}
\cite{usov2000} show that this power ratio is roughly equivalent to $\langle \delta \rangle \simeq 0.1\epsilon_B$, where $\epsilon_B$ is the proportion of energy contained within the magnetic fields. The bolometric radio fluence is related to the observed radio fluence, $\Phi_{\nu}$,  at an observing frequency, $\nu$, for frequencies above the peak radio frequency, $\nu_{\max}$, by:
\begin{eqnarray}
    \Phi_{\nu} = \frac{\beta -1}{\nu_{max}} \Phi_r \left( \frac{\nu}{\nu_{max}} \right)^{-\beta},
\end{eqnarray}
where $\beta$ is the spectral index. Below $\nu_{max}$ there is no observable emission. From \cite{usov2000}, for typical parameters of cosmological GRBs,
\begin{eqnarray}
\nu_{max} \simeq [0.5 - 1]\frac{1}{1+z} \epsilon_B^{\frac{1}{2}} \times 10^6 ~ {\rm Hz}.
\end{eqnarray}
By substitution into Equation \ref{eqn:powerRat}, using the typical value of $\beta = 1.6$ \citep{usov2000} and the observing frequency of 144 MHz, we can show
\begin{eqnarray}
    \langle \delta \rangle & = & \frac{\nu^{\beta}}{\beta -1} \nu_{max}^{1 - \beta} \frac{\Phi_{\nu}} {\Phi_{\gamma}} \\
    \rightarrow \langle \delta \rangle & \simeq & [4.7 - 7.2] \times 10^{9} (1+z)^{0.6} \epsilon_B ^{-0.3} \frac{\Phi_{\nu}} {\Phi_{\gamma}}
\end{eqnarray}
The gamma-ray fluence of GRB 180706A is well constrained, as it was observed by {\it Fermi} GBM to be $(3.3 \pm 0.2) \times 10^{-6}$ erg cm$^{-2}$ in the 10--1000 keV energy band \citep{bissaldi2018}. We can determine a conservative $3\sigma$ 144 MHz radio fluence upper limit, by taking the rms noise in the shortest duration images (multiplying by the duration and 3 to obtain $3\sigma$) from Table \ref{table:rms}. The radio fluence limit at 144 MHz in a 30 second snapshot is therefore $\Phi_{\nu} \le 840 \pm 180$ mJy~s $\lesssim (8.4 \pm 1.8) \times 10^{-18}$  erg cm$^{-2}$ Hz$^{-1}$. Hence, we find
\begin{eqnarray}
    \langle \delta \rangle \lesssim [0.010 - 0.024] (1+z)^{0.6} \epsilon_B ^{-0.3} \label{eqn:delta}
\end{eqnarray}

The redshift is an unknown quantity for GRB 180706A, however we are able to constrain it under the assumption that we would be able to observe coherent radio emission from the prompt emission phase. In addition to the propagation effects within the local environment and host galaxy (outlined in Section \ref{sec:propagation}) and as described in \ref{sec:FRB}, radio emission is subjected to a frequency dependent delay due to plasma along the line of sight, which is commonly referred to as the dispersion delay. We note that, at low radio frequencies, the dispersion delay for extra galactic events may be several minutes (see also Figure \ref{fig:snapshots}). Therefore, if there were prompt radio emission associated with the prompt gamma-ray emission, the radio emission will arrive after the gamma-ray emission. Using the following relation from \cite{taylor1993}:
\begin{eqnarray}
\tau = \frac{{\rm DM}}{241\nu_{\rm GHz}^2} s \label{eqn:dispdelay}
\end{eqnarray}
where $\tau$ is the delay between the emission and the radio signal arriving. Given the 4.5 minute delay between the GRB and the start of the LOFAR observations of GRB 180706A, we would be able to search for prompt coherent radio emission for events with a DM of 1350 pc cm$^{-3}$. Using the approximate relation between DM and redshift \citep[$DM \sim 1200 z$ pc cm$^{-3}$, e.g.][]{ioka2003}, this corresponds to events at $z \gtrsim 1.1$. 

Hence, using the relation $\langle \delta \rangle \lesssim 0.1 \epsilon_B$ and assuming a  redshift of 1.1, we can use Equation \ref{eqn:delta} to constrain the maximum value of $\epsilon_B$ to be $\epsilon_B \lesssim [0.24 - 0.47]$. This value of $\epsilon_B$ is with the expectation for a magnetically dominated GRB jet \citep[e.g.][]{beniamini2014}. Therefore, although we are unable to accurately constrain this model due to the uncertain parameters for GRB 180706A, we show that we are achieving sufficient sensitivity to either confirm a magnetically dominated jet or to rule this out. Using future rapid response radio observations of GRBs we may be able to determine if the jet is baryon dominated or magnetically dominated and answer one of the outstanding questions in GRBs \citep[e.g. ][]{sironi2015}.



\subsection{Constraints on the magnetar central engine model}

In the magnetar central engine model, there are predictions of persistent coherent radio emission from the magnetar. This is in addition to the possible FRB emission considered in Section \ref{sec:FRB}. The persistent emission is typically considered to be from pulsar-type emission \citep[e.g.][]{totani2013}. As we have deep LOFAR observations during the plateau phase, when the magnetar emission is expected to dominate, we can place constraints on this models. First we model the X-ray light curve to determine the magnetar parameters, then we use these parameters to predict the coherent radio emission expected during our observations. 

\subsubsection{Modelling of X-ray light curve}
\label{sec:XRTanalysis}

\begin{figure}
\centering
\includegraphics[width=0.48\textwidth]{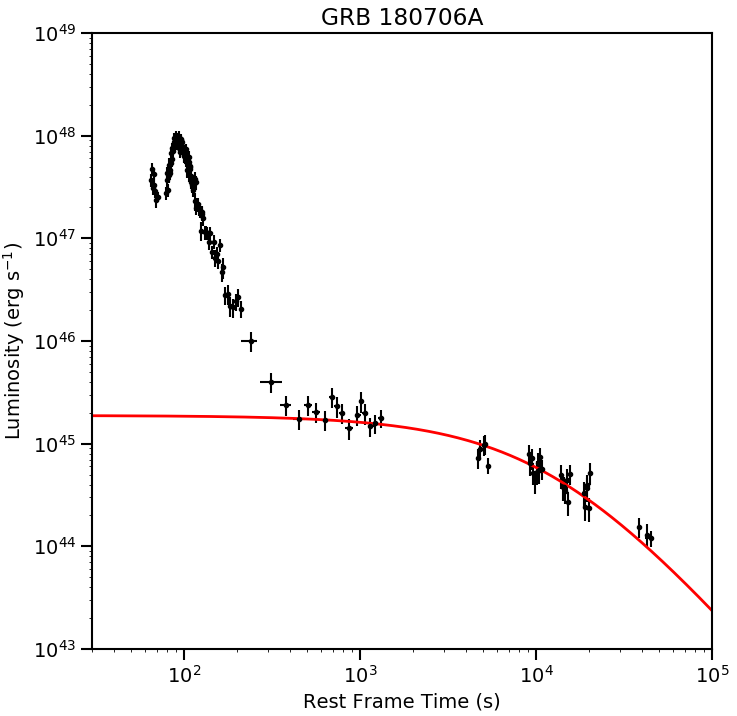}
\caption{This figure shows the rest frame X-ray light curve (black data points, XRT data only), assuming a redshift of 0.2. The red line shows the magnetar model fit obtained, corresponding to a magnetic field of $B = 11.83^{+0.76}_{-0.71} f \times 10^{15}$ G and spin period of $P = 29.37^{+0.68}_{-0.64} f$ ms.}
\label{fig:restframeLC}
\end{figure}

In the case of GRB 180706A, the redshift is unknown so the luminosity and rest frame duration of the X-ray plateau are subsequently unknown. Therefore, in this analysis we assume a redshift value to calculate and fit the rest frame light curve. We choose $z=0.2$ as an arbitrary reference point and show how the results obtained with this chosen redshift can be scaled to other redshift values. Assuming the spectrum can be described by a single power law (as fitted in the X-ray), the light curve was then converted into a rest frame 1--10\,000 keV luminosity light curve using a k-correction \citep{bloom2001} giving an approximately bolometric light curve shown in Figure \ref{fig:restframeLC}.

The rest frame light curve was fitted with the magnetar model \citep{zhang2001}, using the method described in \cite{rowlinson2013}. The magnetar model is given by:
\begin{eqnarray}
B_{15}^2 & = & 4.2025 M_{1.4}^2 R_{6}^{-2}L_{0,49}^{-1}T_{{\rm em}, 3}^{-2} f, \label{eqn:B} \\
P_{-3}^{2} & = & 2.05 M_{1.4}R_{6}^{2}L_{0,49}^{-1}T_{{\rm em}, 3}^{-1}f, \label{eqn:P}
\end{eqnarray}
where $B=10^{15} B_{15}$ G is the magnetic field of the magnetar, $P = 10^{-3} P_{-3}$ s is the initial spin period of the magnetar, $R = 10^{6} R_6$ cm is the radius of the magnetar, $M = 1.4 M_{1.4}$ M$_{\odot}$ is the mass of the magnetar, $T = 10^{3} T_{{\rm em},3}$ is the plateau duration and $L=10^{49} L_{0,49}$ is the plateau luminosity. Here,
\begin{eqnarray}
f=\left(\frac{\epsilon}{1-\cos\theta}\right)^{0.5} \label{eqn:f}
\end{eqnarray}
is a factor encompassing all the uncertainties in the beaming angle, $\theta$, and the efficiency of conversion of the spin energy into X-rays, $\epsilon$. When $f=1$ the system is assumed to emit isotropically with 100 per cent efficiency. Using the observational constraints from \cite{rowlinson2014}, it can be shown that $f \sim 3.45$ by calculating the average of this ratio for all the combinations of beaming angle and efficiency that produced a $>95$\% probability of being able to explain the observed data \citep[c.f. figure 3 in ][]{rowlinson2014}. 

As shown in \cite{rowlinson2019}, the magnetic field and spin parameters can be scaled to different redshifts using these relations:
\begin{eqnarray}
B_{15} & \propto & \frac{(1+z)}{D_{L}}, \label{eqn:Bpropz} \\
P_{-3} & \propto & \frac{(1+z)^{\frac{1}{2}}}{D_{L}}. \label{eqn:Ppropz}
\end{eqnarray}
A cosmology calculator can then be used to convert the required redshift to a luminosity distance ($D_{L}$)\footnote{e.g. the cosmology calculator http://www.astro.ucla.edu/$\sim$wright/CosmoCalc.html \citep{wright2006}}.

We find that the plateau, and subsequent decay phase, can be fitted by a newly-formed stable magnetar with a magnetic field of $11.83^{+0.76}_{-0.71} f \times 10^{15}$ G and spin period of $29.37^{+0.68}_{-0.64} f$ ms assuming a redshift of 0.2. This model is plotted in Figure \ref{fig:restframeLC}. 

\subsubsection{Pulsar like emission}

In the emission model proposed by \cite{totani2013}  \citep[see also][]{pshirkov2010}, coherent emission is expected to occur via magnetic braking (dipole spin-down) of a newly formed magnetar. \cite{totani2013} assumed that this emission is comparable to that of known pulsars and predict that this emission can be described by:
\begin{eqnarray}
F_{\nu} \simeq 8\times10^{7} \nu_{\rm obs}^{-1} \epsilon_{r} D_{\rm lum}^{-2} 
B_{15}^{2} R_6^6 P_{-3}^{-4} ~{\rm Jy} \label{eqn:totaniFlx1}
\end{eqnarray}
where $D_{\rm lum}$ is the luminosity distance in Gpc, $\nu_{obs}$ is the frequency in MHz, $B_{15}=\frac{B}{10^{15}}$, $P_{-3}=\frac{P}{10^{-3}}$, $R_{6} = \frac{R}{10^{6}}$, $R$ is the neutron star radius in metres, and $\epsilon_r$ is the efficiency.

These predictions assume that the pulsar magnetic field axis is highly aligned with the rotation axis of the system to enable the emission to escape via the region that the relativistic jet has cleared. This also ensures that the pulsar emission is directed towards the Earth. As shown in \cite{rowlinson2017}, this assumption is reasonable as the dynamo mechanism produces a magnetic field along the rotation axis and there is insufficient time for the rotation and magnetic axes to become misaligned.

Finally, although the model proposed by \cite{zhang2001} assumes that the newly born magnetar emits via dipole radiation, this is not necessarily the case as the magnetic fields may initially be in a different orientation (e.g. quadrupolar). This means the assumption made by \cite{totani2013}, that the newly born magnetar behaves like the known pulsar population, may be unreasonable. Recently, \cite{lasky2017} investigated this assumption by modelling the late-time decay slope following the plateau phase. They found that the braking indices are consistent with the known pulsar population and have the first detection of a braking index of 3, which is the value expected for pure dipole radiation. Therefore, it is likely reasonable to assume that these new-born neutron stars are spinning down in a similar manner to known pulsars.

In this analysis, we have assumed a redshift of $z=0.2$ for GRB 180706A to allow us to predict the expected flux density of the radio emission at that redshift. However, as shown in \cite{rowlinson2019}, the predicted radio flux density is directly proportional to the observed X-ray fluxes because they originate from the same emission process. Hence, the predicted radio flux density for this event is independent of the actual redshift of the event.

\begin{figure}
\centering
\includegraphics[width=0.48\textwidth]{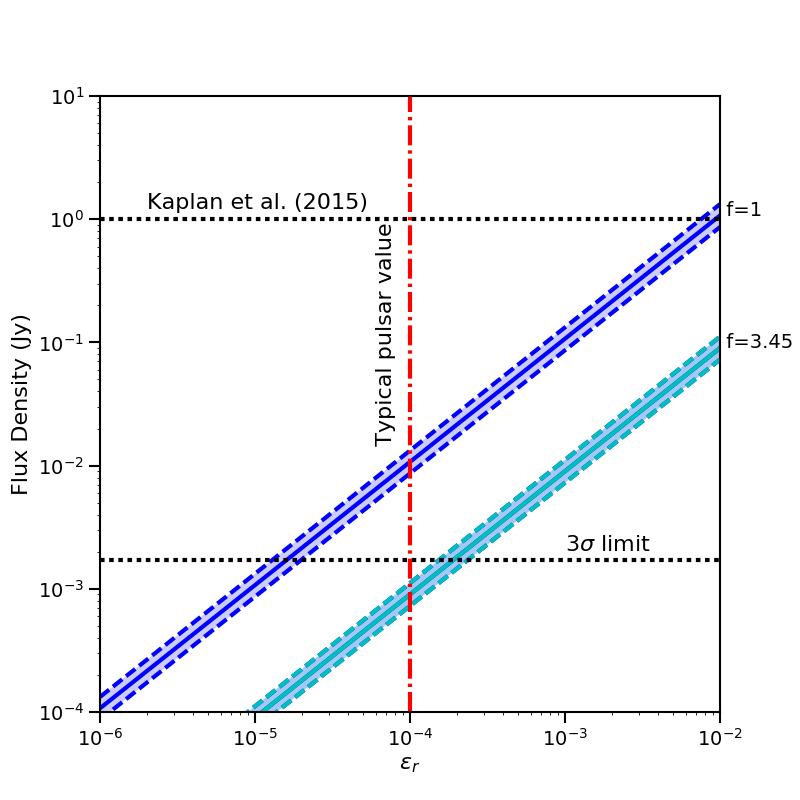}
\caption{This figure shows the predicted flux density at 144 MHz as a function of the efficiency of converting the rotational energy into observable radio emission. The red dash-dotted line shows the typical efficiencies for observed pulsars in our own Galaxy. The lower black dotted line shows the 3$\sigma$ upper limit at the position of the GRB in the 2 hour integrated image. The upper black dotted line shows the previous best limits for this emission from \citet{kaplan2015} for a similar event. The blue line, dashed lines and shaded region illustrates the emission predicted using the parameters obtained for the magnetar from the X-ray plateau assuming 100\% efficiency and isotropic emission (f=1). The equivalent region in cyan represents the more likely scenario with some combination of beaming and reduced efficiency, corresponding to f=3.45.}
\label{fig:totani}
\end{figure}

The efficiency is the remaining unknown quantity in this analysis but, given that the emission is believed to be the same as for pulsars, we use the pulsar value of $10^{-4}$ and illustrate how the predictions vary for a range of values in Figure \ref{fig:totani}. As the plateau phase observed in GRB 180706A fits the magnetar central engine model, our LOFAR observations for the entire duration of the plateau phase are ideal to test for this emission.  The plateau has a long duration, so we can assume that we are in a non-dispersed regime and thus the predicted flux densities are equivalent to the observed values. By using the values fitted for GRB 180706A in Section \ref{sec:XRTanalysis}, an assumed distance of 987 Mpc\footnote{corresponding to a redshift of 0.2 and using the cosmology calculator http://www.astro.ucla.edu/$\sim$wright/CosmoCalc.html \citep{wright2006}}, the mid-frequency of our LOFAR observations (144 MHz), and a 10 km neutron star radius, we find a predicted flux density of
\begin{eqnarray}
F_{\nu} \simeq 10.7^{+2.6}_{-2.1} f^{-2}~{\rm mJy} \label{eqn:totaniFlx2}
\end{eqnarray}

By assuming isotropic emission and 100 per cent X-ray efficiency, we can predict an upper limit of the flux density from a pulsar system to be $\sim 10$ mJy. When the beaming and efficiency factor is assumed to be f = 3.45 \citep[using the analysis of ][]{rowlinson2014}, the predicted flux drops to $\sim 1$ mJy. Our limits are three orders of magnitude deeper than the previous best obtained by the MWA \citep{kaplan2015}. Assuming that the newly formed magnetar is emitting consistently with the known pulsar population and that the emission can escape the system, the emission would have been likely, or close to, detectable in our observations if its beaming and efficiency properties are consistent with the known GRB magnetar population \citep{rowlinson2014}. However, if this system were less efficient and/or more highly beamed than the standard GRB magnetar population, the emission would not be detectable in our observations. Further deep observations of more GRBs with a plateau phase will be required to rule this scenario out.

\section{Conclusions}
\label{sec:conclusions}

In this Paper, we have presented the LOFAR rapid response observations of GRB 180706A, starting at $\sim$ 4.5 minutes of the trigger, searching for coherent low frequency radio emission from the central engine. A detection of coherent radio emission would be strong supporting evidence for the Poynting flux dominated jet or the magnetar central engine model. The X-ray data of GRB 180706A are shown to fit the magnetar model, making it a good candidate to search for this emission.

No emission was detected at the location of GRB 180706A in the full 2 hour integrated image. Neither was emission detected in the short duration snapshot images, which were targeting FRB like emission. We note that the snapshot images were sensitive to particular DM values; an image plane de-dispersion strategy \citep[such as that conducted by ][]{tingay2013} will be required to conduct a deeper search. Future work will include a development of an image plane de-dispersion strategy for LOFAR. Additionally, we plan to introduce commensal imaging and beamformed observations into the LOFAR rapid response mode, enabling us to conduct a standard FRB search.

Due to the 4.5 minute response time and the unknown redshift of GRB 180706A, we are unable to constrain the presence of coherent radio emission associated with a Poynting flux dominated jet. However, we demonstrate that LOFAR is attaining the radio sensitivity required to constrain this model with future GRBs.

The non detection of coherent radio emission associated with the X-ray plateau phase currently does not rule out the magnetar central engine model. This is due to a number of reasons:
\begin{itemize}
   \item The redshift of GRB 180706A is unknown, hence it may be too distant for us to detect FRB like emission.
   \item There remains significant uncertainty in the coherent emission models, ranging from efficiency factors to the spectrum of the emission.
   \item Although the X-ray spectrum implies that GRB 180706A may have occurred in a reasonably low density environment, long GRBs are typically expected to be found in high density environments and hence the coherent emission may not be able to escape.
\end{itemize}

In order to confidently rule out or detect the predicted coherent radio emission, we need multiple rapid response observations of GRBs with radio telescopes of comparable (or better) sensitivity to LOFAR.

\section*{Acknowledgements}
We thank the LOFAR Radio Observatory for implementing the new rapid response mode and for supporting our observations. MJK, AS and RAMJW acknowledge funding from the ERC Advanced Investigator grant no. 247295. WLW and MJH acknowledge support from the UK Science and Technology Facilities Council [ST/M001008/1]

This paper is based (in part) on data obtained with the International LOFAR Telescope (ILT) under project code LC10\_012. LOFAR \citep{vanhaarlem2013} is the Low Frequency Array designed and constructed by ASTRON. It has observing, data processing, and data storage facilities in several countries, that are owned by various parties (each with their own funding sources), and that are collectively operated by the ILT foundation under a joint scientific policy. The ILT resources have benefited from the following recent major funding sources: CNRS-INSU, Observatoire de Paris and Universit{\'e} d'Orl{\`e}ans, France; BMBF, MIWF-NRW, MPG, Germany; Science Foundation Ireland (SFI), Department of Business, Enterprise and Innovation (DBEI), Ireland; NWO, The Netherlands; The Science and Technology Facilities Council, UK; Ministry of Science and Higher Education, Poland.

This work made use of data supplied by the UK {\it Swift} Science Data Centre at the University of Leicester and the {\it Swift} satellite. {\it Swift}, launched in November 2004, is a NASA mission in partnership with the Italian Space Agency and the UK Space Agency. {\it Swift} is managed by NASA Goddard. Penn State University controls science and flight operations from the Mission Operations Center in University Park, Pennsylvania. Los Alamos National Laboratory provides gamma-ray imaging analysis.




\bibliographystyle{mnras}
\bibliography{bibliography.bib} 






\bsp	
\label{lastpage}
\end{document}